\documentstyle[11pt,aaspp4,tighten]{article}
\tighten
\eqsecnum

\begin{document}

\font\twelvei = cmmi10 scaled\magstep1 
       \font\teni = cmmi10 \font\seveni = cmmi7
\font\mbf = cmmib10 scaled\magstep1
       \font\mbfs = cmmib10 \font\mbfss = cmmib10 scaled 833
\font\msybf = cmbsy10 scaled\magstep1
       \font\msybfs = cmbsy10 \font\msybfss = cmbsy10 scaled 833
\textfont1 = \twelvei
       \scriptfont1 = \twelvei \scriptscriptfont1 = \teni
       \def\mit{\fam1 }
\textfont9 = \mbf
       \scriptfont9 = \mbfs \scriptscriptfont9 = \mbfss
       \def\bmit{\fam9 }
\textfont10 = \msybf
       \scriptfont10 = \msybfs \scriptscriptfont10 = \msybfss
       \def\bmsy{\fam10 }

\def\ea{{\it et al.}~}
\def\etal{{\it et al.~}}
\def\eg{{\it e.g.,~}}
\def\ie{{\it i.e.,~}}

\def \msol {\rm{M}$_\odot$}
\def \mdot {\rm{M}$_\odot$~yr$^{-1}$}
\def \kms{km~$\rm{s}^{-1}$}
\def \cc{$\rm{cm}^{-3}$}
\def \arcs{\char'175}
\def \lam{$\lambda$}
\def \micra{$\mu$m}
\def\lsim{\raise0.3ex\hbox{$<$}\kern-0.75em{\lower0.65ex\hbox{$\sim$}}}
\def\gsim{\raise0.3ex\hbox{$>$}\kern-0.75em{\lower0.65ex\hbox{$\sim$}}}

\title{Simulating Electron Transport and Synchrotron Emission in Radio Galaxies:\\
	Shock Acceleration and Synchrotron Aging in
	Axis-symmetric Flows\altaffilmark{4}}

\author{T. W. Jones\altaffilmark{1},
        Dongsu Ryu\altaffilmark{2}
       and Andrew Engel\altaffilmark{1,3}}

\altaffiltext{1}{Department of Astronomy, University of Minnesota,
    Minneapolis, MN 55455: twj@msi.umn.edu}
\altaffiltext{2}{Department of Astronomy \& Space Science, Chungnam National
    University, Daejeon 305-764, Korea:\\ryu@canopus.chungnam.ac.kr}
\altaffiltext{3}{Department of Physics, University of Arizona
    Tucson, AZ 85721: aaengel@physics.arizona.edu}
\altaffiltext{4}{Accepted for publication in the Astrophysical Journal}

\begin{abstract} 
We introduce a simple and economical but effective method for 
including relativistic electron transport in multi-dimensional
simulations of radio galaxies. The method is designed to follow
explicitly diffusive acceleration at shocks, and, in smooth flows,
second-order Fermi acceleration, plus adiabatic and synchrotron
losses for electrons in the energy range responsible for radio emission
in these objects. We are able to follow both the spatial and energy (momentum)
distributions of the electrons, so that direct synchrotron emission
properties can be modeled in time dependent simulated flows of this type
for the first time.
That feature is essential if simulations are to bridge successfully the
fundamental physical gap between flow dynamics and observed emissions.

As an initial step towards that goal, 
we present results from some
axis-symmetric MHD simulations  of Mach 20 light jet flows.
These explicitly explore the effects of shock acceleration, as well
as adiabatic expansion and synchrotron aging in smooth flows.
The simulations demonstrate the importance of the
fact that even for steady inflows
jet terminal shocks are not simple, steady
plane structures. Most importantly this should play a very major role 
in  determining the
properties of synchrotron emission within the terminal hot spot
and in the lobes generated by the jet back flow. In fact, the outflows
are inherently complex, because of
the basic ``driven'' character of a jet flow.
Consequently, the nonthermal
electron population emerging from the jet may encounter a wide range
of shock types and strengths, as well as magnetic field environments.

We may expect to find
a complex range in synchrotron spectral and brightness patterns associated
with terminal hot spots and lobes. These include the possibility of
steep spectral gradients (of either sign) within hot spots, the potential 
in lobes for islands of flat spectrum electrons within steeper spectral 
regions (or the reverse) and spectral gradients coming from
the dynamical history of a given flow element rather than from
synchrotron aging of the embedded electrons.
Finally, synchrotron ``aging'' in the lobes tends to proceed more 
slowly than one
would estimate from regions of high emissivity. That is a consequence
of the fact that those regions are ordinarily places where the
magnetic fields are the strongest, so that the instantaneous
rates of energy loss are atypical of the full history of the
electron population. This feature supports earlier suggestions
that nonuniform field structures may help to explain why dynamical
ages of FRII sources often seem to be greater than the apparent age of 
the electrons radiating in the lobes, as measured in terms of
spectral steepening, or absence thereof.
\end{abstract}


\clearpage

\section{Introduction} 

The standard paradigm for radio galaxies (RGs) is
based on high speed plasma jets, formed in active galaxy
nuclei (AGN), but penetrating far into circumgalactic environments and
creating giant lobes of luminous material as a consequence
of interactions with the intergalactic medium (\eg \cite{brid92}).
The defining radio emission represents synchrotron radiation from
relativistic electrons and magnetic fields. Those two crucial constituents
are thought to be transported from the AGN, enhanced  or generated 
by the jet-IGM encounter or both. 
Modern radio interferometry has provided richly detailed
intensity, spectral and polarimetric images of RGs (\eg \cite{leahy97}), while rapidly
advancing computational tools have allowed increasingly sophisticated
multi-dimensional magnetohydrodynamical (MHD) simulations of the plasma 
flows (\eg \cite{clarke96}).
This paper presents the first such simulations that explicitly include 
time dependent transport of the relativistic electrons that produce the observed 
radio emission in such objects.
\cite{cb88} carried out an early related calculation by mapping an
analytic electron distribution onto a simulated bow-shock flow pattern,
while \cite{jk93} followed the time dependent behavior of a 
simplified electron distribution during the evolution of a shocked
gas cloud.
\cite{clarke89} and \cite{ms90} computed early models of synchrotron
emission from RGs based on simulated fluid dynamical variables and intelligent
guesses at the possible relativistic electron properties.
But, no published simulations have followed the full electron
distribution through a time-dependent evolution in a simulated RG-like
flow.
That important feature is a very difficult technical challenge, because the
length and time scales needed to model effectively the microphysics of
electron transport are typically many orders of magnitude smaller than
those convenient to include in full scale hydrodynamical or magnetohydrodynamical
simulations of RGs.

To avoid these difficulties we introduce here a simple, economical 
method of relativistic
electron transport that directly utilizes those mismatched scales 
and is suitable for time-dependent RG simulations. 
The method is designed to include the effects of diffusive
shock acceleration and second-order Fermi acceleration, as well as adiabatic 
compression or expansion
and synchrotron radiative cooling of the electrons in smooth flows.
We present results of its initial application.
There are many dynamical effects that are likely to
influence the eventual
synchrotron brightness distribution and spectrum to be expected
in realistic models of RGs. Before it makes sense to attempt anything
that could be termed a ``real'' model of a RG, it is crucial that
the individual dynamical influences be understood. The present paper,
therefore, is intended to begin addressing these individual factors.
The role of shock acceleration of RG electrons is centrally important (\eg
\cite{heav87}; \cite{bland87}; \cite{tak89}; \cite{krulls92})
While there are many studies of the physics of particle acceleration at
plane or spherical shocks (\eg \cite{kj91}), there are few previous
simulations designed to explore shock acceleration in complex flows (but see
\eg \cite{jkt94}) and, as mentioned, none for flows similar to those
expected in RGs.
Therefore, our first priority is to explore how shock acceleration
may be characterized and recognized in such environments, or the degree
to which such flows may tend to confuse simple interpretations of this
process.

Given these objectives and the fact that
2D, axis-symmetric flows are much easier to understand than
3D flows and still provide a first cut at the properties of
RG flows, we limit this initial study to that problem. We have begun
an extension of this work into fully 3D simulations and will present
those results elsewhere.
Since synchrotron cooling (commonly called ``aging'') is also likely to be a significant
effect and will compete with and mask the influence of shock 
acceleration,
we also include here an example of a flow that includes both of these
effects together. We defer to later papers inclusion of second-order
Fermi acceleration, not because it is necessarily unimportant, but because
it is significantly more difficult to establish a unique, physical
characterization of the momentum diffusion controlling it. Thus, at
this stage, it would add significant complexity and uncertainty to an already
complex set of questions.

The plan of the paper is this. In \S 2 we outline our methods, including
the new electron transport scheme. Section 3
introduces the parameters of the jet flows we have modeled, while \S 4
discusses our results. A brief summary of key findings is given in \S 5.
An appendix contains a detailed discussion of practical requirements
for numerical treatment of electron transport in RGs and justification
for the scheme we introduce here.

\section{Methods} 

\subsection{Dynamics}

We evolve the equations of ideal nonrelativistic magnetohydrodynamics
(MHD) in cylindrical coordinates $(r,\phi,z)$, with $\phi$ an ignorable
coordinate. All three components of velocity and magnetic fields
are included, so the model is $2\frac{1}{2}$-D. The code is an MHD
extension of Harten's (\cite{harten83}) conservative, second-order finite difference
``Total Variation Diminishing'' scheme, as detailed in \cite{ryuj95};
\cite{ryujf95}; \cite{ryuyc95} and \cite{rmjf98}.
The code preserves $\nabla\cdot B = 0$
at each time step using an approach similar to the Constrained Transport (CT)
scheme (\cite{eh88}) as described in detail by \cite{ryu98}.
We use a passive ``mass fraction'' or ``color'' tracer, $C_j$, to
distinguish material entering the grid through the jet orifice 
($C_j = 1$), or from ambient plasma ($C_j = 0$).

\subsection{Electron Transport}

Our electron transport scheme is an adaptation of the standard
diffusion-convection equation 
for charged particles (\eg \cite{ski75}); 
\begin{equation}
{\partial f \over \partial t } = {1 \over 3} p {\partial f \over \partial p}(\nabla
 \cdot \vec u) - \vec u\cdot\nabla f + \nabla \cdot (\kappa \nabla f) +
 \frac{1}{p^2} \frac{\partial}{\partial p}\left ( p^2 D \frac{\partial f}{\partial p}\right ) + Q,
\label{dceq}
\end{equation}
where $f(\vec x, p, t)$ is the isotropic part of the
nonthermal electron distribution, $\kappa$ is
the spatial diffusion coefficient, $D$ is the momentum diffusion
coefficient, $Q$ is a source term that
represents the net effects of ``injection''
and radiative losses at a given momentum,
$p$, while the thermal plasma velocity is $\vec u$.
This equation is valid for ``fast'' (superthermal) particles when scattering is 
strong enough to keep the particle distribution almost isotropic and 
is appropriate on lengths large compared to the particle scattering 
length itself.

The first term on the right accounts
for adiabatic compression or rarefaction in the background flow.
Eq. [\ref{dceq}] is not valid within shocks, but when integrated across 
a velocity {\it discontinuity}, the first three 
terms account for ``first order Fermi acceleration'' at shocks, 
also known as ``diffusive shock acceleration''.
The fourth term allows for second-order Fermi acceleration resulting
from particle interaction with Alfv\'enic turbulence (\eg \cite{ski75};
\cite{bland87}). 

As mentioned in the introduction and explained in detail in
the Appendix, there is a severe mismatch between RG dynamical scales and
diffusive transport scales that apply to eq. [\ref{dceq}] for 
electron momenta relevant to
radio synchrotron emission. That mismatch makes it impractical with
conventional computational methods to solve eq. [\ref{dceq}] as it
stands to study electron transport in full RG flows. 
That is, the electron diffusive lengths and
times are vastly smaller than those appropriate to the gasdynamics.
The key physical fact is that
the characteristic energy of electrons radiating synchrotron emission can be
stated roughly as $E \sim 3~\nu^{\frac{1}{2}}_{9}/B^{\frac{1}{2}}_{10}$ GeV,
where $\nu_{9}$ is the observed frequency in GHz, and
$B_{10}$ is the magnetic field compared to $10~\mu$G or nT.
Thus, the electrons of interest mostly have energies $\lsim 10$ GeV
($p \lsim 10^4$ mc).
The gyroradius of such particles is $r_g \sim 3\times 10^{11} E_{GeV}/B_{10}$ cm.
This leads to characteristic diffusion lengths and shock
acceleration times in RGs $\lsim$ AU and $\lsim$ yr, respectively 
compared to flow scales typically measured in kpc and kyr or
larger. The most serious consequence of this comes
from the requirement that numerical shocks must appear thinner than
a particle diffusion length for accurate solutions of eq. [\ref{dceq}] 
(see the Appendix). 

However, as also explained in the
Appendix, that same mismatch can be exploited to develop a 
simplified equation of electron transport, provided 
$f(p)$ is sufficiently broad that it can be represented as a piecewise
power law over finite momentum bins.  
This last feature is very natural, in fact, in light of the same
small time and length scales for electrons, since very rapid
diffusive shock acceleration
guarantees that GeV electrons emerge from shocks 
with power law momentum distributions on time scales much less than
time steps required by the MHD.
For much higher electron energies these simplifying conditions
break down, but the method we derive could
be used to provide a ``low energy injection spectrum'' for those particles,
as well.

For continuity with the remainder of our discussion we present here
the simplified transport equation, but refer readers to the Appendix
for derivation details and tests of its validity.
We divide the momentum range of 
interest into a modest number, N, of logarithmically spaced bins 
bounded by $p_0,\ldots,p_N$.
Defining $y_i = \ln{p_i/p_0}$, bin widths can be expressed as, 
$\Delta y_i = y_{i+1} - y_i = \ln{p_{i+1}/p_i}$.
We can integrate equation [\ref{dceq}] within each momentum
bin to define $n_i = 4 \pi \int_{p_i}^{p_{i+1}} p^3 f d\ln{p}$ as the number of
electrons in the bin.
For convenience we normalize $n_i$ by the total plasma mass density, to
form $b_i = n_i/\rho$, then
assume a piecewise power law
$f(p) = f_i~ (p/p_i)^{-q_i}$  within $p_i \le p \le p_{i+1}$, so that
$n_i$ is given in terms of $f_i$, $q_i$ and $\Delta y_i$ by eq. [\ref{nieq}].
For these relatively low energies diffusive transport at shock 
discontinuities is properly handled
by defining the form of the electron distribution just down
stream of a shock to be the steady state power law appropriate to the jump
conditions for that shock. That fixes the
ratio $b_{i+1}/b_i$ at shocks.
Electrons are injected from the thermal plasma at shocks using a 
common injection model, so that a fixed fraction, $\epsilon$, of the
total electron flux through a shock is injected and accelerated to the
appropriate power law momentum distribution.
Away from shocks, in smooth flows, but where the diffusion lengths
of the electrons are much smaller than dynamical lengths, spatial
diffusion is negligible.
We can easily include the effects of synchrotron 
``aging''.
When all of these features are included eq. [\ref{dceq}] becomes (see
also  eq. [\ref{nieq}])
\begin{equation}
{{d b_i}\over{d t}} = 4 \pi
\left (\frac{1}{3}\nabla u
-\frac{q D}{p^2}
+ \frac{1}{\tau_{so}}\frac{p}{\hat p}\right )
\frac{p^3 f} {\rho}
\left\vert_{_{p_{i}}}^{^{p_{i+1}}}\right.,
\label{simplet}
\end{equation}
where $\tau_{so} = \frac{3}{4}
\frac{(m c)^2}{\sigma_T U_B}\frac{1}{\hat p}$ defines the 
synchrotron cooling time at 
momentum $\hat p$, which we take arbitrarily to be $\hat p = 10^4$ mc.
In this expression, $\sigma_T$ is the Thomson cross section, 
and $U_B = B^2/(8\pi)$.
Note that the right hand side of eq. [\ref{simplet}] is the difference
between fluxes at the two momentum boundaries of the bin $i$. Thus, we have
a conservative, finite volume scheme that depends on a
simple model of sub-grid structure (in momentum space) for its accuracy.

Although straightforward to include, we now drop the second order Fermi 
term, ``$D$'' from these initial simulations for the following reasons. 
First, in the vicinity of
shocks, this acceleration process is likely to be much less
efficient than first-order Fermi acceleration. That is
simply because the second-order process comes from nearly balanced
energy fluxes of waves propagating parallel and antiparallel to
the magnetic field that resonantly scatter with the electrons; \ie
on ``isotropic'' Alfv\'enic turbulence.
First-order Fermi acceleration, on the other hand, depends only on
waves propagating in the streaming direction that 
will be generated by the streaming particles themselves.
In either situation resonant waves have wavelengths comparable
to the particle gyroradius.
Downstream of the shocks where particle streaming may be
less important, it is significant for such low particle
energies that the resonant wavelengths are probably within a couple orders
of magnitude of thermal ion gyroradii (taking $u_s \sim 10^8$ cm s$^{-1}$).
Such waves may be fairly strongly dissipated, for example,
by nonlinear Landau damping (\eg \cite{leevolk73}).
Thus the level of relevant isotropic Alfv\'enic turbulence will
likely depend on a local cascade of hydrodynamical turbulence (\eg \cite{grap82}).
While we will certainly want to understand what role post-shock,
second-order acceleration might play (\eg \cite{eil86}; \cite{krulls92}),
only ad hoc models could be
applied at present. Since this set of computations
represents the first effort to treat nonthermal electron transport
inside time dependent models of RGs,  and since there are
many possible competing influences to be understood before
``realistic'' models are sensible,
the best initial strategy is to restrict ourselves to
the most straightforward physics that is likely to be important. 
First-order acceleration is efficient and almost certain to happen at shocks, so
its role must be understood from the beginning.

\subsection{Computation of Synchrotron Emissivity}

From the spatial distribution of $b_i$ and $q_i$ along with
the MHD variables, $\rho$ and $B$, it is straightforward to compute
the synchrotron spectral emissivity, $j_\nu$. 
Ignoring corrections due to slow curvature in the electron
spectrum, $j_\nu$, can be expressed as  (cf. \cite{jos74})
\begin{equation}
j_\nu = j_{\alpha o} \frac{4\pi e^2}{c} f(p) p^q \left (\frac{\nu_{B_\perp}}{\nu}\right ) ^\alpha \nu_{B_\perp},
\label{emiss}
\end{equation}
with $\alpha = \frac{q - 3}{2}$,
$j_{\alpha o} \sim 1$ (cf. \cite{jos74}) and 
$\nu_{B_{\perp}}=\nu_B \cos{\theta}$ the electron
cyclotron frequency in terms
of the magnetic field projected over the angle $\theta$ onto the plane of 
the sky.
$f(p)$ along with $q$ are found from mapping $b_i$ and $q_i$ onto
the critical synchrotron frequency as
$p = m c \sqrt{\frac{2 \nu}{3 \nu_{B_{\perp}}}}$.
We have interpolated $q$ between momentum bin centers to give $\alpha$
a continuous form.
Eq. [\ref{emiss}] would break down for strongly curved spectra, but
by that point in calculations of the type used here the local index will be
too steep for the emission to be significant.
In a calculation designed to predict the surface brightness distribution
of a RG model one should include explicitly the effects of variations in
$\cos{\theta}$. That is not our intention here, and such a calculation
would need to evaluate carefully the effects of ``sub-grid'' fluctuations
in the field directions and of the orientation of the jet with respect
to the line of sight. Our aim here is to understand how electron  
acceleration and transport are likely to be reflected in the overall
emission properties of a region within a RG, independent of the location
of the observer. For those
purposes it is better to ignore the factor $\cos{\theta}$, so 
will in this paper simply take it to be unity. 
Again, we are {\it not} in this paper
computing surface brightness
distributions from these simulations, which would
involve line-of-sight integrations of $j_\nu$. 
That would be inappropriate, given
the axis-symmetric nature of the simulations, and would 
further exaggerate the dynamical constraints imposed by this symmetry.

\section{The Simulated Jet Properties}

Our initial simulated  MHD jets are all dynamically identical.
They have a simple,  ``top hat'' velocity profile, are Mach 20 
with respect to the uniform ambient medium ($M_j = u_j/c_a = 20$), in
thermal pressure balance with it, and have a density contrast
$\eta = \rho_j/\rho_a = 0.1$.
The jet enters at $z = 0$, with an initial radius of 36 zones, while the
entire uniform grid is $384\times 1536$ zones ($10\frac{2}{3} r_j \times 42\frac{2}{3} r_j$).
Defining length and time in units of initial jet radius ($r_j = 1$) and ambient
sound speed ($c_a = \sqrt{\gamma P_a/\rho_a} = 1$, with $\gamma = \frac{5}{3}$), 
the simulations are followed 
for about 11.7 time units, when
the bow shock of the jet reaches the right $z$ boundary.
Reflecting boundaries are used along the jet
axis, while continuous boundaries are used elsewhere.
There is a background poloidal magnetic field $(B_{background})$
($B_\phi = B_r = 0$; $B_z = B_{zo}$), with
a magnetic pressure 1\% the gas pressure; 
\ie $P_b = \frac{1}{\beta} P$, with $\beta = 10^2$.
The in-flowing jet carries an additional toroidal magnetic field component
appropriate to a uniform axial jet current with a return current
on the surface of the jet; \ie $B_\phi = 2\times B_{zo} (r/r_j)$
for $r \le r_j$. At $r = r_j, \beta = 20$. The in-flowing jets are 
slightly over-pressured at the outside.

These simulations are truly MHD rather than HD, despite the apparent weakness
of the initial magnetic field.
Even at the initial strengths modeled here
the magnetic fields play a role in the evolution, especially in the cocoon.
There are numerous locations where the plasma $\beta \sim 10$, so
that its pressure is not entirely ignorable. There is, however, a
more important and much less recognized role for the field
in flows initialized with even $\beta >> 1$. In 2-D and 3-D MHD
simulations of the Kelvin-Helmholtz instability, for example, 
we have demonstrated crucial dynamical contributions from
magnetic fields in flows with initial $\beta > 10^3$ (\eg \cite{jgrf97};
\cite{rjf98}; \cite{jrf98}). The point is
that through field ``stretching'', magnetic tension can be increased in 
complex flows to the
point where it is comparable to or exceeds the Reynolds stresses
of the gasdynamical flow ( coming from spatial variations in $ \rho u_iu_j$), even though the
magnetic pressure may be smaller than the gas pressure. The relative
importance of Maxwell to Reynolds stresses can be roughly represented
by the Alfv\'en Mach number in a flow, $M_A = u/v_A$.
When the local $M_A$ drops to small values in a complex flow, that
can lead through reconnection to so-called ``dynamical alignment'' and  flow ``self-organization''.
There are also many locations within our model flows where $M_A \lsim 1$.
In short, the flows are made smoother by the presence of the field, even
though in the original configuration the field nominally appeared too weak to
influence the dynamics.
A comment may be in order on the meaning of magnetic reconnection
obtained from an ``ideal MHD'' code. While reconnection is most
fundamentally a topological transition (\eg, Axford 1984), it cannot
take place without resistive, dissipative effects on small scales. In an ideal MHD
code these dissipative effects are numerical in origin, so we cannot
model the microphysics of driven reconnection (which is still not understood
at this time, in any case). However, there is evidence that the phenomenology of
reconnection is often correctly captured. For example, Miniati \etal (1998),
using the Cartesian version of the code we employ here,
demonstrate clear development in 2-D supersonic MHD cloud simulations 
of the so-called ``resistive tearing-mode''
instability that is physically associated with driven reconnection.

We present here three examples of the electron transport within the
dynamics of the above jet flows. Their properties are listed in Table \ref{tabmod}.
Electrons are modeled in the momentum range $p_0 = m_e c$ and $p_N \approx 1.63\times 10^5 m_e c$. 
All three include the effects of diffusive shock acceleration. {\bf Models 1} and
{\bf 2} are alike in that the electrons have negligible influence from 
synchrotron aging , while {\bf Models 1} and {\bf 3} are alike in
that the electron populations originate entirely with the
in-flowing jet. {\bf Model 2} differs in that the electron population is
mostly injected locally from thermal plasma at shocks within the 
modeled flow. {\bf Model 3}
differs, then, in that electrons are influenced very significantly by
synchrotron aging.

\begin{deluxetable}{cccccccc}
\footnotesize
\tablecolumns{6}
\tablecaption{Summary of  Simulations \label{tabmod}}
\tablehead{ \colhead{}& \colhead{} & \colhead{} & \colhead{} & \colhead{}
&\colhead{}&\colhead{} &\colhead{} \\
\cline{3-4}\\
\colhead{Model \tablenotemark{a}} &
\colhead{N\tablenotemark{b}} &
\colhead{In-flowing Electrons\tablenotemark{c}}&
\colhead{Jet Spectrum\tablenotemark{d}} &
\colhead{Shock Injection\tablenotemark{e}} &
\colhead{Cooling Time\tablenotemark{f}} &
\colhead{$B_{background}$\tablenotemark{g}}\\
\colhead{}&\colhead{}&\colhead{($b_1$)}&\colhead{$q_{jet}$ ($\alpha_{jet}$)}&
\colhead{($\epsilon$)}&\colhead{($\tau_{so}$)}&\colhead{$\mu$Gauss}
}
\startdata
1 & 4 & $10^{-4}$ & 4.4 (0.7) & $0.0$ & 40 & 10 \nl
2 & 4 &  $10^{-6}$ & 4.4 (0.7) & $10^{-4}$& 40 & 10 \nl
3 & 8 & $10^{-4}$ & 4.4 (0.7) & $0.0$ & 4 & 30 \nl
\enddata
\tablenotetext{a}{All models used Mach 20 jets ($M_j = u_j/c_a = 20$), 
with a density
contrast, $\eta = \rho_j/\rho_a = 0.1$. Units come from a jet radius, $r_j = 1$, 
an ambient density, $\rho_a = 1$, and a background sound speed,
$c_a = \sqrt{\gamma P_a/\rho_a} = 1$ ($\gamma = 5/3)$.
The cylindrical computational box was 10.67 $r_j$ units high and 42.67 $r_j$ long.
End time for each simulation was $t = 10.67$ units.
There was a uniform, poloidal jet magnetic field, $B_{z0}$ that also
extended to the background, so $B_{z0} =  B_{background}$ with $\beta = P/P_b = 100$, while 
the incoming
jet also contained a toroidal field, $B_{\phi} = 2\times B_{z0}(r/r_j)$.}
\tablenotetext{b}{Number of electron logarithmic momentum bins spanning the
range $p_0 = 1$ mc to $p_N = 1.63\times 10^5$ mc. ($\ln{p_N/p_0} = 12$)}

\tablenotetext{c}{Ratio of nonthermal to thermal electrons in the
incident jet flow.}
\tablenotetext{d}{$q_{jet}$ is the momentum distribution power law index of
the in-flowing electron population; $\alpha_{jet}$ is the corresponding
synchrotron index.}

\tablenotetext{e}{Assumed fraction of the thermal electron population
injected to the nonthermal population as it passes through a shock.}

\tablenotetext{f}{Time for electrons to cool below momentum, $\hat p = 10^4$ mc 
in the ambient, or background, magnetic field, measured in time units $r_j/c_a$.}

\tablenotetext{g}{Field that would produce the listed cooling time,
assuming illustrative values for the simulated jet: $u_j = 0.1 c$, $r_j = 1$ kpc.
This would also map the displayed synchrotron emissivity data onto 1.4 GHz.
Applying the additional constraints on the jet listed above, the implied jet
density and kinetic luminosity are 
$\rho_j \approx 3\times 10^{-29} B^2_{background}$ g cm$^{-3}$ and 
$K_{kinetic} = \frac{\pi}{2} \rho_j u^3_j r^2_j 
\approx 1.2\times 10^{43} B^2_{background}$ erg sec$^{-1}$, 
with $B_{background}$ in $\mu$Gauss.}

\end{deluxetable}

In the ``adiabatic'' {\bf Models 1} and {\bf 2}, there is very little
curvature to the electron spectra, so in those models we used four
momentum bins (\ie $N = 4$) giving
$\ln{\frac{p_{i+1}}{p_i}} = 3$. (See the Appendix for justification of these
numbers.)
In {\bf Model 2} we injected electrons at $p_0$ using the simple model
described in \S A.3.
Electrons should be injected at slightly suprathermal postshock
energies, so the value $p_0 = m_e c$ was selected as appropriate for jets
with speed $u_{j} \sim 0.1 c$, assuming that shocked electrons
are thermalized to a temperature comparable to but smaller than the
ions.
Since the nonthermal, cosmic-ray electrons are passive, all results could be scaled for different
choices of $p_0$.

To include synchrotron aging it is necessary to relate the characteristic
cooling time $\tau_{so}$ defined in association with eq. [\ref{simple}]
to the computational time units, here given by $r_j/c_a$.
The cooling time depends as well on the magnetic field strength
and the momentum $p$ as
\begin{equation}
\tau_{s} = 2.5\times10^1 \frac{1}{ p_4}~
\frac{u_{j8}}{M_j}
\frac{1}{r_{jk}} \frac{1}{B_{10}^2},
\label{tcool}
\end{equation}
where $\tau_s = \tau_{so}\frac{\hat p}{p}$, $p_4$ is the electron
momentum in units $10^4$ mc,
$u_{j8}$ is the jet speed expressed in units of $10^8$ cm s$^{-1}$,
$M_j$ is the jet Mach number (here set to $M_j = 20$), $r_{jk}$
is the incoming jet radius in kpc, and $B_{10}$ is
the strength of the magnetic field in units of 10 $\mu$G,
or equivalently in units nT.
By comparison, the end points of our simulations are $t = 10\frac{2}{3}$.

For {\bf Models 1} and {\bf 2} we have set $\tau_{so} = 40$, the
objective being to produce negligible aging for electrons
of interest during the simulation. To project emission from
electrons with 
momenta $p \lsim 10^4$ mc into the radio band, we should be considering
magnetic fields $\gsim 10~\mu$G. Assuming, for
example, that $B_{z0} = 10~\mu$G,
and $u_{j8} = 30$, then $r_{jk} \approx 1.1$, so
that the length of our computational box would be about 45 kpc, and
the time unit, $r_j/c_a \approx 7\times 10^5$ years.

By contrast, for {\bf Model 3} we set $\tau_{so} = 4.0$, so that
synchrotron aging is now very significant  for electrons in the
energy range of interest over the time scale of
the simulation. Keeping the
same physical jet speed and radius, as well as the time unit,
this case would correspond to an axial magnetic field $B_{z0} \approx 30~\mu$G.
Other associated dependent parameters are outlined in Table 1.
These combinations are illustrative only and not meant to be particularly
representative of any real RGs. Other combinations would work
as well for our present purposes, which are aimed at the physics of
the electron transport. 
For the simulation of {\bf Model 3}, we used eight momentum
bins ($N = 8$) to cover the same range as the
other models, as justified in \S A.4.

\section{Discussion} 

\subsection{Flow Dynamics}

We first outline some key dynamical properties
of the simulations, remembering that the MHD properties of the different
electron transport models are identical except for dimensional scaling
factors. 
Fig. 1 shows at $t = 10.67$ images of the plasma density, $\rho$, the
magnetic field intensity, $B$, and gas compression, 
$\frac{d \ln{\rho}}{d t} = -\nabla\cdot \vec u$.
The compression highlights shock structures.
These images remind us that such flows are complex and unsteady.
All of the structures are ephemeral or at least highly variable. 
Jets, as strongly driven flows are {\it not} reasonable to model as
equilibrium structures (cf. {\cite{sato96}).
As described by many before us (\eg \cite{norm85}; \cite{Lind89}),
once equilibrium is broken, jet plasma
alternately expands and then refocuses while interacting with its
self-generated cocoon, creating oblique jet
shocks as a result. 
\footnote{In this paper ``oblique'' and ``perpendicular'' refer to the angle between the
shock face and the velocity field, not the shock normal and the
magnetic field, as is customary in the particle acceleration literature.}
Those shocks are neither steady nor stationary,
however. Most important to our discussion, the oblique jet shocks interact episodically
with the terminal jet shock, causing it to vary in both strength and structure.
The jet never re-establishes an equilibrium in its flow.
The terminal shock also includes
a ``Mach stem'', so that in 2-D at least, jet flow coming down the outside of the
jet, near $r_j$, always
exits through an oblique shock. At times there is
little or no perpendicular terminal shock;
then, most of the emerging jet flow is only weakly shocked.
When a perpendicular portion to the
terminal shock exists, it usually is strong, with a
compression ratio $\sigma \approx 3.8$ ($\rightarrow q \approx 4.08;
\alpha \approx 0.54$) in these simulations. The many other shocks and the
oblique segments of the terminal shock are mostly weak (Fig. 1c), but 
they can exhibit density jumps, $\sigma \ga 2 - 3$ ($\rightarrow q 
\approx 6 - \frac{9}{2}$).
Along the jet axis intersecting oblique shocks can be strong, but
these intersections intercept a small fraction of the flow, so will not
produce a large population of flat spectrum electrons.

Vorticity is shed from the outer edge of the terminal shock
to form the turbulent jet cocoon.
There are distinct episodes of strong ``vortex shedding'' coincident
with disruption and reformation of the terminal shock.
Once shed, the vortices interact with the Kelvin-Helmholtz unstable
boundary layer of the jet, generating an even more complex back-flow
and perturbing the jet flow, as well.
All of these dynamical features are represented in 
the Fig. 1. snapshot.
The large ``rolls'' visible in Fig. 1 are remnants of vortex rings that were shed
earlier than the time displayed. 
The magnetic field in the back-flow is dominated within an
axis-symmetric flow by $B_\phi$, 
both because this magnetic component is enhanced by stretching in the
expanding flow emerging from the jet,
and because magnetic reconnection of the $B_r$, $B_z$ field
components annihilates most of that field from the back-flow in an
axis-symmetric geometry. This flux annihilation inside vortical
flows is well-known and sometimes termed ``flux expulsion'' 
in the MHD literature (\cite{weiss}).
In a 3-D flow both the vortex and magnetic field
structures would be stretched, twisted and tangled;
\ie the back-flow would become turbulent  and disordered
on small scales (\eg \cite{norman96}; \cite{rjf98}; \cite{jrf98}) leading to vortex 
and magnetic flux tube complexes. 

\subsection{Electron Transport \& Emissivity}

The electron transport we model takes place ``on top of'' this dynamics;
\ie, the electrons do not feed back dynamically on the flow.
To check this for consistency we confirmed that for the numbers 
we assume, the electrons never represent
more than about $0.1$ \% of the kinetic energy density in the jet or
the back flow.
Remember that the dynamics behind the electron transport and
synchrotron simulations discussed below is identical in each model.
The electron distribution, and especially its momentum dependence,
reflect the flow history of a given fluid element as 
appropriate for the particular transport assumptions, as well as the
current fluid condition. Our objective in this paper is to begin to learn how
to interpret the emergent patterns. Since the astronomical tool for this
is synchrotron emission, we will explore the physics in that paradigm using
local emissivities.
We expect that this approach will make much more tractable the
interpretation of full 3D simulations.
For the convenience of the discussion we express emissivities
as they would be at $1.4$ GHz, based on the fiducial magnetic field
values listed at the end of \S 3 and in Table \ref{tabmod}.
They can be easily rescaled by choosing other combinations of
source parameters.
Illustrative examples of the emissivity properties in these 
models are shown in Fig. 2~-~4.
Fig. 5 explores the correlations between emissivity, spectral index
and magnetic field strength using the same data displayed in the
other figures.
At the start we point out the obvious ``limb brightening'' of the
emissivities seen in all of the models displayed.
This feature is not characteristic of real radio lobe surface brightness
in FRII sources, but is an artifact of the assumed
axis-symmetry of the simulations; particularly the persistence of the
large ``rolls'' in the back-flow and the decoupling that occurs in
axis-symmetry between $B_{\phi}$ and ($B_r$, $B_z$).
In an equivalent 3-D simulation the bright structures should be better
mixed into the turbulent interior regions of the back-flow.
This is one reason we defer any display of model surface brightness
until we have 3-D structures in the models.
That does not detract from the present discussion, however, since
this initial exploration is mostly aimed at identifying physical
relationships in emission properties and flow properties, as well as
recognizing emissivity characteristics that we may eventually be able
to use as diagnostics of the history of the local electron population.

\subsubsection{``Adiabatic'' {\bf Models 1} and {\bf 2}}

Fig. 2a and 2b provide grayscale images of $\log{j_\nu}$
at 1.4 GHz for the ``adiabatic'' electron  models {\bf 1} and {\bf 2}
at $t = 10.67$.
For these two models emissivity values from eq. [\ref{emiss}] are 
displayed when they are within a factor  $3\times 10^3$ of the
peak emissivity.
Smaller emissivities are blacked out.
For comparison we also display with the same dynamic
range in Fig. 2c a ``pseudo emissivity'',
based on the MHD properties of the flow; \ie $j_c = C_j P B^{\frac{3}{2}}$, 
modeled after the approach introduced in \cite{clarke89}.
We have added the jet color variable, $C_j$ to Clarke's original
definition, because of differences in our assumed background magnetic
field. Clarke's magnetic field vanished outside the jet flow and
its cocoon,
whereas we used a simpler model that continues the axial field into the
background. Since $C_j = 0$ in the background
our definition allows $j_c > 0$ only for regions 
containing jet plasma, so that they are comparable to Clarke's.
On the other hand, for the {\bf Model 1}, {\bf 2} and {\bf 3}
emissivities based explicitly on transported relativistic electrons, 
the absence of a significant electron population
in the background makes this factor immaterial.
Fig. 3a and 3b show  with an inverse gray scale the 1.4 GHz spectral
index distributions, $\alpha = - \partial \log{j_\nu}/\partial \log{\nu}|_{1.4} = (q-3)/2$, ($\alpha_1$ and $\alpha_2$) over the same
emissivity range in {\bf Models 1} and {\bf2}.

Fig. 2 compares emissivities produced by the 
two ``adiabatic'' models represented by $j_1$ and $j_2$.
to the ``pseudo emissivity'', $j_c$.
The three emissivity images have clear similarities.
All show a dominant ``hot spot'', down stream (to the 
right) of the jet terminal shock. 
Each produces qualitatively similar patterns of bright emission 
within the back-flow.
The jet is illuminated in the $j_c$ model emissivity with patterns
that resemble those of the $j_1$ model.
Of course, the jet is largely invisible in $j_2$, because the
jet electron population is assumed very small in that model.
The model congruencies reflect the strong roles of two key ingredients
to the emissivity; namely, the strength of the magnetic field and
adiabatic compression and expansion.
Note, however, that $j_1$ and $j_2$
have much greater local contrast than $j_c$.
Since the magnetic field distributions are identical for
all three emissivity calculations, differences in the
effective radiating electron distributions must be responsible.
That is apparent by noting within smooth flows that
$n_i \propto \rho^{\frac{q_i}{3}}$,
so $j_{1,2} \propto n_i B^{\frac{q_i - 1}{2}} 
\propto  \rho^{\frac{q_i}{3}}  B^{\frac{q_i - 1}{2}}$. This is actually
rather similar to $j_c \propto \rho^{5/3} B^{3/2}$, except for 
variations modeled in $b_i = n_i/\rho$
and $q_i$, but not modeled by $j_c$.
The formula given for $j_c$ is based on an assumed $\alpha = 0.5$, so
contrast could be increased in $j_c$ by assuming a significantly
steeper index. The natural extension of $j_c$ to general 
$\alpha$ would be $j_c  \propto  p^{2\alpha} \rho ^{1-2\alpha} 
(B sin\psi)^{\alpha+1} \nu^{-\alpha}$ (D. Clarke, private communication).
For $\alpha = 1$  this gives $j_c \propto (P B)^2/\rho$, just as for
$j_{1,2}$ with $\alpha = 1$.
That would
not represent the same effect noticeable in $j_1$ and $j_2$, however,
since from Figs. 2, 3 and 5, we can see that large $j_{1,2}$ come primarily from
regions where $\alpha < 0.7$.

\subsubsection{``Synchrotron Aged'' {\bf Model 3}}

In Fig. 4 we show the analogous simulated synchrotron emissivity and spectral
information for our
{\bf Model 3}, which includes the effects of synchrotron aging . Recall
that, except for this influence, {\bf Model 3} is identical to {\bf Model 1}.
Thus, Fig. 4a ($j_3$) can be compared to Fig. 2a ($j_1$), while Fig. 4b ($\alpha_3$)
can be compared to
Fig. 3a ($\alpha_1$). In addition, we illustrate the level of
spectral curvature  of
{\bf Model 3} in Fig. 4c,
using of the difference between the simulated spectral indices at
1.4 GHz and 5.0 GHz ($\delta_3 = \alpha_3(5 GHz) - \alpha_3(1.4 GHz)$). 
Note that we use formal definitions of $\alpha$ rather than  the
traditional observed value defined
by the ratio of fluxes measured at two distinct frequencies.
The intrinsic ``observed'' index can be found by the simple formula
$\alpha^{1.4}_5 = \alpha + 0.5*\delta_3$.
These ``observed'' indices would be almost the same as $\alpha$ near the origin
of the jet and in flows just either side of the terminal shock in
Fig. 4. On the other hand the ``observed'' spectra would tend to be $\sim 0.05$
steeper in mid regions of the jet and in most of the back flow, cocoon
regions bright enough to show in this figure.
But, we mention  another caveat with regard to comparisons with
directly observed properties. Beyond other complications already mention,
including differences between emissivity and surface brightness, a real
observation would likely smooth over a physically significant range of sight lines.
Non-black regions in Fig. 4 cover a 
wider range of emissivity values (namely a factor $3\times 10^4$) than in 
Figs. 3 and 4, since the inclusion of synchrotron aging adds considerable
contrast to the emissivity distribution. This factor 10 extension captures approximately
the same physical regions as for the other models.

The emissivity distribution, $j_3$, for {\bf Model 3} including
synchrotron aging, is shown in Fig. 4a.
As mentioned earlier, the dynamical range found in $j_3$ is greater
than for any of the other models, since there is considerable
steepening visible in the spectral index as one moves away from
electron sources; namely, the jet origin and strong shocks. 
Steepening due to aging is apparent
even within the jet, where $\alpha_3$ increases from $0.7$ at the 
origin to values near $1$ even before the first oblique shocks are
encountered.  Although the emissivity distribution within the jet in
Fig. 4a is very similar in appearance to that for {\bf Model 1} in
Fig. 2a, the factor 10 increase in dynamic range covered in Fig. 2a slightly
de-emphasizes the stronger emissivity decreases caused by synchrotron aging.
By contrast to the adiabatic models, where the
synchrotron spectral index remains near $\alpha \approx 0.7$ almost 
everywhere in
the jet, the oblique jet shocks do produce a noticeable flattening in the
spectra of {\bf Model 3}. The oblique shocks have compression ratios,
$\sigma \lsim 3$ corresponding to a postshock
index, $q > 4.5; \alpha > 0.75$, so they only 
flatten an incident spectrum that has been steepened by radiative
losses. In fact, from Fig. 4c it is apparent that in {\bf Model 3}
there is significant spectral curvature  in the
synchrotron spectrum even within the jet. 
The nature of that is also visible in the electron momentum distribution
plot in Fig. 7c.
While the oblique jet shocks are relatively ineffective as electron 
accelerators, places where these oblique shocks intersect form strong
shocks with their normals aligned to the jet axis. They produce
spectra synchrotron spectra as flat as $\alpha \sim 0.52$ (see
Fig. 5). But, as already mentioned, the
portion of the jet flow intercepted by these shocks is very small,
so in these simulations they have little impact on the large scale
synchrotron emission within these flows.

\subsubsection{General Comments -- Electron Transport \& RG Dynamics}

We begin this section with some observations about patterns in the
synchrotron emissivity spectra within the jet structures themselves. 
In Fig. 3 we see that the
jet spectra for the adiabatic {\bf Model 1} and {\bf 2} are almost
everywhere $\alpha = 0.7$, which was the injected spectrum.
That is despite the existence of several jet shocks  and apparent ``knots''
in the flow
before the terminal shock, where the spectrum flattens to $\alpha \approx 0.54$.
Should we have expected to see spectral flattening at the jet ``knots''?
In these two models the answer is no, since shocks only flatten an
incident electron spectrum if it is steeper than the index $q_s = 3\sigma/(\sigma - 1)$,
where $\sigma$ is the compression ratio through the shock (see \S\S 2.2 and A.3.2). 
Thus, only shocks with compression ratios,
$\sigma > 3.14$  would lead to flattening within these jets.
In fact, except very near the jet axis the
jet shocks are oblique, with compression ratios, $\sigma < 3$,in these simulations.
There is one very restricted flatter spectrum region in the jet in
{\bf Models 1} and {\bf 2} right on the axis
and just to the left of the terminal shock (Fig. 3). There a very
strong shock has formed at the intersection of oblique shocks.
However, as mentioned earlier, the volume of jet plasma involved is very small, so
this feature is hardly noticeable.
On the other hand, synchrotron cooling in {\bf Model 3} steepens
the incident electron spectra coming into the ``knots'', so that
in Fig. 4b we can see clear evidence for spectral flattening at the 
jet ``knots'', albeit to indices, $\alpha \gsim 0.7$, in these simulations.
The details of such features will probably vary with the details of
the assumed jet structures and Mach numbers. For example, if the
jet flows and magnetic fields conspired to produce the greatest
emissivities close to the jet axis, then spectral changes at
shocks would be more apparent, since jet shocks do tend to be transverse
on axis.

The observational evidence bearing on this is still limited.
Our choice of $q_o = 4.4~(\alpha_o = 0.7)$
was intended to represent a ``typical'' FR II RG jet spectrum, of course.
It is most often hard to isolate structures in RGs that are
clearly ``only'' jet flows (see, \eg \cite{rud96}) because jets are
often relatively weak emitters and because they can, therefore, be confused by
contamination from the cocoon surrounding them. This makes it especially
hard to isolate changes from the jets to the knots, because the
contamination from cocoon emission (expected to be relatively steep-spectrum)
is greater for the inter-knot regions than in the knots. 
In some cases where the jet data look convincing,
such as for M87 (actually a FR I source), the radio spectrum seems to be 
virtually constant
along the entire jet with no indication of curvature (\eg \cite{bir91}; \cite{meis96}).
The extension of this spectrum to the visual band is what strongly
constrains aging in M87, since the estimated cooling times for
radio emitting electrons are longer than typical estimates of
propagation times.
And, indeed, there are indications, despite the constancy of
$\alpha$ in the radio band, that the radio-to-optical spectrum
steepens between the ``knots'' in the M87 jet
(\cite{sparks96}).
In Cyg A the jet radio spectrum appears also to be
locally nearly straight, but to exhibit different slopes at
different locations (\cite{rud96}; \cite{katz93}).
Other radio jets may show evidence of steepening of the spectrum along the jet 
(\eg \cite{mack98}), but data do not yet allow detailed analysis
in terms of electron transport. Clarke \etal 1986 reported evidence
for spectral flattening  at a knot in the jet of  Cen A. The 
spectral index flattens to $\alpha \approx 0.7$, from the surrounding
emission where $\alpha \approx 1.5$. If these changes really are 
within the jet flow and not influenced by contamination, this case 
could easily be accounted for by the effects we see in our simulations.

The spectral index distributions in Fig. 3 show for our adiabatic
models that
much of the bright emission outside the jet itself, especially in the 
hot spot, is associated with 
spectra flatter than the $\alpha = 0.7$ expected from
the in-flowing jet plasma.
Even the brightest parts of the cocoon are dominated by spectral indices with $\alpha \lsim 0.7$.
That must be largely due to acceleration in
the terminal shock, of course. Closer examination reveals considerable 
fine structure, however, in the spectra. Fig. 5 shows, especially for
{\bf Model 2}, that reasonably bright emission is produced with
spectral indices ranging roughly over a wide range, $\alpha \sim 0.54 - 0.75$.
The $\alpha \approx 0.54$ emission comes from electrons 
accelerated in the strong, plane portions of the terminal shock.  
Near the
head of the jet in this model the brightest emission in the back-flow still
tends to have
relatively flat spectra, $\alpha \la 0.6$. However, it is
interesting to note that the spectra of
the brighter areas in {\bf Models 1} and {\bf 2}
are often steeper farther back in the cocoon. 
That does not come from radiative losses, which are negligible in these two
models,
but from increased mixing between flat and steep electron populations.
This is an important effect to recognize, since it mimics what is 
usually assumed to be due to synchrotron aging in the back flow.
The specific locations of flat and steep spectra in these examples, should
be viewed with caution, of course, since once again we are observing axis-symmetric
flows that do not mix in the same ways as the more general 3-D flows will do. 

Where do these steep-spectrum electrons come from in the adiabatic models?
The steeper spectrum emission in the hot spot 
and the cocoon represents plasma that has escaped
passage through the strong, plane terminal shock. 
One such path is plainly visible for {\bf Model 1} in
Fig. 3a. There we can trace emission with $\alpha = 0.7$ all the way
from the jet origin on the left into the cocoon back-flow. There
are even $\alpha = 0.7$ ``fingers'' of unmodified jet
electron populations just outside the jet and near
the left of the grid. The source of these electrons is
made apparent by the invisibly low $j_2$ values (Fig. 2b) in the same regions.
That plasma  flows into the cocoon 
through the outer, oblique portion of the terminal
shock where the shock compression
$\sigma < 3$, so the spectrum is unmodified from its original
form in the absence of synchrotron
aging.
We note in an axis-symmetric flow that plasma emerging from the
jet terminus near the outer radius of the jet must initially
flow back into the cocoon closer to the jet channel than plasma
emerging along the jet axis, since 2D
flow ``streamlines'' do not cross.  That also explains why near the
hot spot flatter spectrum emission tends to extend
along the outer, forward parts of the cocoon in {\bf Models 1} and {\bf 2}, 
since that represents plasma passing through a strong terminal shock
nearer the jet axis.
As noted, turbulence within the back-flow eventually 
mixes these populations together, steepening the average spectrum. 
There is also a related very important source of steep spectrum
electrons in the cocoon in these adiabatic models;
\ie there are time periods with little or no perpendicular terminal
shock. Then most or all of the emergent jet plasma 
carries electron populations unflattened by the terminal shock. 
This is especially relevant when there is fresh electron injection at
shocks, as in {\bf Model 2}.
There, the steeper spectrum emissivities can be introduced by electron injection
in oblique portions of the terminal shock.

Finally, in the context of the adiabatic models we mention that
there is
significant, bright emission in {\bf Model 2} coming from electrons
in the shocked, ambient IGM; \ie where $C_j = 0$, so $j_c = 0$. 
This comes from IGM
electrons injected to the relativistic population by the jet bow shock and now 
embedded in regions with moderately strong magnetic fields.
Fig. 3b shows that the spectral range is large in these regions,
with $\alpha \ga 0.6$. 

At first glance the emissivity distribution for the
synchrotron aged {\bf Model 3} in Fig. 4a
is very similar to the adiabatic models, especially {\bf Model 1}, which
shares the same electron injection history. Even accounting for the
greater dynamic range displayed in $j_3$ this is especially true near
the terminal hot spot. Within the hot spot that similarity extends to the
spectra, which are relatively flat in both models. But, closer examination
reveals some interesting differences that come from the effects of
strong synchrotron aging. First, even within the hot spot, {\bf Model 3}
displays a substantial spectral range, from close to $\alpha \sim 0.5$
to values greater than $\alpha \sim 1$. In fact, the brightest regions of
the hot spot have spectral indices just greater than $\alpha = 0.7$,
by contrast to values near 0.54 for {\bf Models 1} and {\bf 2} (see
Fig. 5). 
The strong spectral steepening that takes place at the very front of
the terminal hot spot region is a product of the very strong magnetic
field that forms at the nose of the emerging jet.
This is evident in Fig. 4e and 4f, where it is apparent that the
emission from the strongest fields (which lie in this region),
while very bright, is also quite steep.
From Fig. 5 we can see that the strongest fields in the hot spot
are $\sim 8$ times greater than $B_{background}$, so from eq. [\ref{tcool}],
the cooling times of the emitting electrons are quite short; namely,
$\tau_s \sim 6\times 10^{-2}r_j/c_a$. In that situation, with the
absence of any additional acceleration within the postshock flow,
spectral steepening is a virtual certainty.
In fact, strong spectral index gradients in some RG hot spots have been
identified using ``spectral tomography'' techniques (\cite{tr98}),
showing that real hot spots have structures that may eventually
enable us to explore electron radiative histories and origins, once we 
understand the behaviors of more realistic models.

For the most part the brighter regions in the cocoon of
{\bf Model 3} have spectra with $\alpha \gsim 1$, and on the
whole, the cocoon has a steeper spectrum than the hot spot. These
properties are in rough, general
agreement with observational expectations (\eg \cite{denn97}; \cite{mack98}). 
It is interesting to note that there are
``pockets'' of flat spectrum emission in the back flow, however.
There is a particularly prominent region in the forward-most swirl
just above the terminal hot spot, visible in Fig 5. Electrons in these 
regions were
subjected to strong shock acceleration, but were shed in a vortex
event without ever passing through a region of strong magnetic field
capable of causing strong aging.
The field in their current position is actually below $B_{background}$
(see
Fig. 1b), so their cooling time has become a large fraction 
of the full simulation time. These regions are moderately bright because
the electron density is relatively high.
As possible evidence that such structures might form in real RGs
we mention that
\cite{mack98} point to the existence of ``flat spectrum
islands''($\alpha \sim 0.6$)  in the lobes of NGC 315 where 
$\alpha \sim 1.3$.

The presence of flatter spectrum populations within
the back flow is also related to the fact that additional steepening of
the spectrum of the cocoon 
is slight in the brightest regions 
in {\bf Model 3} once one looks past the
most forward portions (see Fig. 4b). From Fig. 1b and Fig. 5f
it is plain that the magnetic fields in the visible regions farther back in the
cocoon are roughly comparable to those in the forward
cocoon or the much of the terminal hot spot. In particular they
are greater than $B_{background}$. Thus, the nominal
cooling times based on current conditions would be $\tau_s < 4 r_j/c_a$,
while the age of the jet at this time is $t \approx 10.67 r_j/c_a$.
The reason for this apparent contradiction between short cooling times
and slow aging is the highly
unsteady and nonuniform nature of the cocoon flow, especially the
magnetic field, in the cocoon. Because of that the
electrons spend only a fraction of their lives in strong
field regions where they emit strongly. Thus, they cool much more 
slowly than
one would naively predict, based on emissivities. 
This supports previous suggestions that
the radiative ages of RGs appear less than their dynamical
ages because of
small magnetic field filling factors (\eg \cite{myer85})
or filamentary magnetic field structures (\eg \cite{emw97}).

We can recognize some additional and generally
interesting behaviors from Fig. 5.
For example, from Fig. 5a and 5c for most of the
points representing shocked jet emission (including all the
points with $\alpha < 0.7$ in Fig. 5a) there is an
envelop restricting the strongest emission to the
flattest spectra. Note, for example, in {\bf Model 1} none of the brightest
emission outside the jet comes from regions with $\alpha > 0.6$, despite
the fact that the
magnetic fields in the fainter, steep spectrum emitting regions
are as strong as in the bright, flat spectrum regions (see Fig. 5b).
The ``nose'' of the $(j_\nu, \alpha)$ distribution in Fig. 5a, 5c, 5e
corresponds to the terminal hot spot emission, where fields are strong {\it and}
electron densities are high.
The largest emissivities
naturally correspond to the strongest field regions. The converse
is not true, however; strong fields do not necessarily
produce a large emissivity.
The other ingredient in $j_\nu$, of course, is
the electron density, $f(p)$. 
Fig. 5 is, therefore, showing us a wide range of
electron densities associated with magnetic field strengths contributing
to the emission,
and even some degree of anti-correlation between field and $n_i$ in the back-flow.
Back-flowing plasma has mostly undergone
adiabatic expansion after emerging from the jet, while the $B_\phi$
component of the magnetic field has been enhanced in some locations
by stretching of
field lines at the same time.
In {\bf Model 3} (represented in Fig. 5e,f), where synchrotron aging
also acts to reduce emission by steepening the spectrum,
the patterns are actually
very similar to those in the adiabatic models. However, much of the emission
is shifted upward to steeper spectra; roughly by $\Delta \alpha \sim 0.2$.
The points remaining near $\alpha \sim 0.5$ correspond to shocked points
within the jet near its axis, and points just past the jet
terminal shock, where aging effects are not yet important.
Finally, we comment that more generally, there are 
many ``micro patterns'' visible in Fig. 5. 
Those tend to trace hydrodynamical features, so that they are revealing 
histories of
the electrons. Unfortunately, those flow patterns are usually unsteady
so that it would be difficult, if not impossible to use them to
determine explicit connections between individual points and flow dynamics.

\section{Conclusions}

We have developed a simple but effective numerical scheme for
transport of low energy relativistic electrons suitable for
studies in time dependent simulations of radio galaxies. The scheme can
follow the evolution of electrons accelerated by first order
Fermi acceleration in shocks, second order Fermi acceleration,
adiabatic cooling and
synchrotron cooling, a.k.a. ``aging'', in smooth flows. From the
electron spatial and momentum distribution and the magnetic field
properties found through the fluid evolution one can directly
compute the radio synchrotron emission.
In this paper we have applied those methods to axis-symmetric
MHD jet flows including first order Fermi shock acceleration and
synchrotron aging effects on the electrons, in order to identify some of 
the behaviors that may be expected in multi-dimensional flows.
We begin with 2D flows since they are much simpler than 3D flows to
study and to understand. 
We have focussed our discussion on effects that should occur in both
2D and 3D flows, so that what we learn here will facilitate future
studies with more general flows. Briefly, the most important findings of
this work that we believe to be independent of our specific
simulation assumptions are:

$\bullet$ The fact that jet terminal shocks are not simple, steady
plane structures should play a major role in  determining the
properties of synchrotron emission within the terminal hot spot
and in the lobes generated by the jet back flow. In fact, the outflows
are inherently complex, because of
the basic ``driven'' character of a jet flow.
As pointed out recently by \cite{sato96}, Complexity
is basic to all driven non-equilibrium plasma flows. RGs are dramatic
examples of such flows, so they should be influenced by episodic patterns of
instability and reorganization. In particular, the strength and form of 
the terminal
shock is constantly and sometimes dramatically changing. 
Thus, the nonthermal
electron population emerging from the jet may encounter a wide range
of shock types and strengths, as well as magnetic field environments.

$\bullet$ Because of the previously mentioned behaviors, we may expect to find
a complex range in synchrotron spectral and brightness patterns associated
with terminal hot spots and lobes. These include the possibility of
steep spectral gradients (of either sign) within hot spots, the potential 
in lobes for islands of flat spectrum electrons within steeper spectral 
regions (or the reverse) and spectral gradients that result from
the dynamical history of a given flow element rather than from
synchrotron aging of the embedded electrons.

$\bullet$ Synchrotron ``aging'' in the lobes tends to proceed more 
slowly than one
would estimate from regions of high emissivity. That is a consequence
of the fact that those regions are generally places where the
magnetic fields are the strongest, so that the instantaneous
rates of energy loss are atypical of the entire history of the
electron population. This feature is apparent even in axis-symmetric
flows; it should be stronger in real, 3D flows, since for equivalent
initial structures, local magnetic field enhancements should be
greater in 3D (\cite{jrf98}). Our finding supports earlier suggestions
that nonuniform field structures may help to explain why dynamical
ages of FRII sources often seem to be greater than the apparent age of 
the electrons radiating in the lobes (\eg \cite{emw97}).

There are many uncertainties inherent in attempts to model such complex
objects as RGs. The flows are inherently intricate and many key parameters
are largely unknown at present. That is why we have examined first the
simplest systems that may be qualitatively representative and have tried 
to restrict
our questions to those not wholly dependent on uncertainties coming from
the limitations imposed by our various choices. 
Even from these few initial results we can see 
the importance of modeling electron acceleration and transport 
and the significance of shock history
to the radio spectra in RG hot spots and lobes. 
The interpretation of synchrotron spectra in RGs clearly is not
simple, but represents the combined influence of
several dynamically coupled factors. 
That should, perhaps point us to another 
analysis approach; namely, thinking directly in terms of ``Complexity'' itself.
\cite{sato96} argue that in such systems
we need to shift our paradigm from a focus on specific elements
(\eg the existence of a terminal shock that can accelerate particles)
to the consequences of interactions among elements (\eg long range
order and the relationships between dynamical structures).
The task then becomes one of understanding how the ``microphysics''
(\eg particle transport) can help to establish relationships among
the various system elements.

\acknowledgments

Support for this work was provided by NSF grants AST93-18959, 
INT95-11654 and AST96-16964, by NASA grant NAG5-5055 and the 
University of Minnesota Supercomputing Institute.  
The work by DR was supported in part by
KOSEF through the 1997 Korea-US Cooperative Science Program 
975-0200-006-2. We
gratefully acknowledge helpful discussions with B.I. Jun  and
especially Hyesung Kang during
development of the electron transport scheme and with Larry Rudnick
about practical connections between simulations and radio galaxies.
We also thank the referee, David Clarke, for very helpful comments
and discussion on the manuscript.

\appendix

\section{Appendix}

\subsection{Constraints on Computational Diffusive Electron Transport in RGs}

A number of studies have considered time dependent solutions to
eq [\ref{dceq}] for charged particles in shocked flows (\eg \cite{fall87}; \cite{drury91};
\cite{kj91}; \cite{kjr92}; \cite{ber95}; \cite{kj95}; \cite{kj97}). 
Those studies and most others like them have focussed on the transport
of nonthermal ions.
Although diffusive transport is distinct for nonrelativistic
electrons (\eg \cite{lev96}), relativistic electrons  can be treated
exactly like ions of similar rigidity, so the above treatments are directly relevant to
the transport of electrons responsible for synchrotron emission.
All of the papers mentioned, however, considered 1-D flows for a single shock system. 
This limited application reflects a severe numerical constraint 
in solving eq. [\ref{dceq}],
especially for low energy  particles.
With existing numerical schemes it is not practical to carry out a
full and direct solution to eq. [\ref{dceq}] within a ``large-scale'',
complex multi-dimensional flow.
The reasons are straightforward.
{\it First, note that eq. [\ref{dceq}] applies only outside of shocks, 
since its validity
is restricted to particles whose scattering lengths and gyroradii
are large compared to the shock thickness, which should be
a few thermal ion gyroradii} (\eg \cite{kj95}; \cite{kj97} and
references therein).
{\it Its application to diffusive shock acceleration depends entirely
on being able to match upstream and downstream solutions properly at
the shock} 
(\eg \cite{drury83}; \cite{drury91}).
An accurate numerical solution to eq. [\ref{dceq}] requires that the 
shock be a {\it discontinuity} or close to it.
Numerical schemes for compressible fluid dynamics, including
the one we employ, spread shock structures over at least one or two zones.
One of the necessary matching conditions for diffusive electron
transport is that
the distribution $f$ be continuous across the discontinuous
physical shock. So, if we apply eq. [\ref{dceq}] at a shock, 
we must demand, to avoid errors within the (unphysical)
numerical shock, that the
advective flux term in eq. [\ref{dceq}] 
integrate to the same value as at a true discontinuity.
It is simple to show that the errors in this are first order in 
$\Delta x_s/x_d$, where $\Delta x_s \ge \Delta x$ is
the numerical shock thickness, $\Delta x$ is a zone width at the shock
and $x_d = \kappa/u_s$ is the 
diffusion length for energetic particles at the shock, with $u_s$ being the velocity
jump across the shock. Therefore, we must have $\Delta x_s/x_d << 1$.

To see what constraint this gives in a RG application we need to
estimate $x_d$.
There is empirical evidence that near shocks the diffusion
coefficient, $\kappa$, is a steeply increasing function of momentum
with a form resembling the so-called Bohm diffusion coefficient, 
$\kappa \sim r_g w$, where $w$ is the particle speed (\eg \cite{egbs93}).
For GeV electrons in a magnetic field exceeding $\sim 10~\mu$G, for
example, we would estimate for Bohm diffusion near a strong 
shock with speed, $u_s \sim 10^8$ cm s$^{-1}$, that  $\kappa \lsim 10^{22}$ cm$^2$ s$^{-1}$ 
and $x_d \lsim 10^{14}$ cm.
Lower energy electrons, especially those near shock injection energy,
would give smaller $x_d$ by at least two orders of magnitude.
Thus, a direct simulation in a RG of the full eq. [\ref{dceq}] for electrons 
even at GeV energies would require numerical shocks much thinner than
$10^{14}$ cm. To include electrons down to
injection energies in this situation
we would necessarily require numerical shocks to be no thicker 
than about an AU.
{\it This is a spatial constraint, so temporal sub-cycling techniques
to deal with associated rapid acceleration are not sufficient to deal with this problem.}
Since the simulation scale in a RG is typically tens or hundreds of kpc 
one would need to resolve shocks on a scale no more than about $10^{-6}$
of the main grid scale and much smaller than this to
capture the particle behavior with eq. [\ref{dceq}] near injection energies. 
For simple, 1-D flows this kind of refinement near shocks may be feasible
using a nonuniform grid, although one still must deal with 
numerical issues coming from the
accompanying orders of magnitude range in the time steps
required for the solution.
An adaptive mesh refinement (AMR) approach may eventually
make the analogous multi-dimensional treatment feasible, but
until those codes are available for MHD, including diffusive
transport, an alternative way around the problem is necessary
for meaningful calculations.

\subsection{A Practical Approach to Dealing with Electron Transport}

Fortunately, for the problem at hand we can actually exploit
the small diffusion length constraint to introduce a simplified version of 
eq. [\ref{dceq}] that still correctly captures the essential physics of
electron transport, at least when the dynamical feedback of the
electrons can be ignored; that is when the electrons can be
seen as ``test particles.''
\cite{jk93} applied a very simple
version of such a scheme several years ago, and \cite{jj98}
have briefly described a variant of the one discussed below.
To begin, we remind readers of the point in \S A.1
that we should distinguish  between
behaviors at shocks and within smooth flows, away from shocks. 
Therefore, we divide discussion of our proposed scheme into
these topics and begin with shocks, since the emergent properties of the
electron distribution help to determine our approach in smooth flows.

\subsubsection{Transport Through Shocks}

At shocks we must properly balance the change
in advection at the shock against diffusion away from it.
From the arguments in the preceding section, we must find a way
to make the numerical shock much thinner than a conventional numerical
zone width.
The most obvious numerical solution to this problem is to 
treat the shock as a real discontinuity; \ie confine
the shock transition to a zone boundary as far as particle acceleration
is concerned. 
Generally, as mentioned 
earlier, that involves imposing the matching conditions to $f$ across the
shock  (\eg \cite{drury83}; \cite{drury91}) to properly
account for both diffusion and advection on both sides of the shock.
\cite{drury91}
used this method to find analytic time dependent test particle
solutions for $f$ at a plane shock, for example. \cite{ber95} used it
numerically to treat time dependent ion acceleration at the
spherical blast wave of a supernova remnant, including backreaction
on the local flow.
Those particular methods are not practical to apply to numerical
schemes in general flow patterns, because they are computationally
intensive and require 
highly customized assumptions.
However, for the low energy electrons in RGs that are of interest to us in 
this paper, it is possible to accomplish the required outcome in a manner
that is easily adapted to general numerical schemes.
The key points are: 1) as already discussed, diffusion lengths
are very small compared to typical numerical zones on either side of
the shock for the particles under consideration; \ie $\Delta x >> x_d$;
and 2) the characteristic diffusion timescale, $t_d = x_d/u_s$ adjacent to
the shock is much smaller than numerical time steps
associated with the standard CFL condition; \ie $\Delta t \sim \Delta x/u_s >> t_d$.
In fact, $\Delta t/t_d \sim \Delta x/x_d$, so, from our earlier
discussion, both conditions are easily satisfied by many orders of
magnitude in RG simulations.

The first point means that the
entire upstream ``precursor'' to the shock that controls
evolution of $f$ is well-contained 
within a single numerical zone. This prohibits treatment of
backreaction by the nonthermal particles on the flow, since that
depends on knowing the velocity gradient within the precursor. But, 
electrons  accelerated in shocks are generally thought not to produce
significant backreaction (\eg \cite{bland87}).
In the work we present here the nonthermal electron population is
not dynamically coupled back to the flow.
The second point also means that the precursor is formed instantaneously
by comparison to the numerical, dynamical time step. More directly, it means
that the timescale for particles to be accelerated at the shock
to the energies of interest is also very small compared to the
numerical time step. Assuming for simplicity that $\kappa$ is
the same upstream and downstream of a strong shock, the mean time for
particles to be accelerated from an initial
momentum, $p_o$, to momentum, $p$,  can be written
approximately as $t_a \approx \frac{20}{u^2_s}~\int_{p_o}^p \kappa(p') d\ln p'$
(\cite{lag83}).
For Bohm diffusion of relativistic electrons near a strong shock
$t_a \sim 20\times t_d \sim 2\times 10^7 E_{GeV}/(B_{10} u^2_{s8})$ sec, 
where $u_{s8}$ is the shock speed expressed
in units of $10^8$ cm s$^{-1}$.
For GeV electrons this is about a year for our selected characteristic
numbers, compared to
a typical CFL-determined dynamical time step in a RG simulation $\gsim 10^3$ years.
The synchrotron radiative cooling, or ``aging''  timescale, 
$t_s \sim 10^{15}/(\nu^\frac{1}{2}_9 B^\frac{3}{2}_{10})$ sec, should
be comfortably longer than either of these for the particles of immediate
focus, so we can deal with effects of radiative cooling
separately on large scales.
It has been demonstrated analytically (\cite{drury91})
and numerically (\cite{kj91}) that after a time $\Delta t$ a particle
distribution $f_o$ flowing into a steady plane shock emerges downstream 
with the appropriate steady state solution 
for $p/p_{at} << 1$, where $p_{at}$ is the momentum
that can be reached by diffusive acceleration during the time interval $\Delta t$,
as defined by $t_a$.
For our application, where $p_{at}(\Delta t)$ is much larger
than momenta of interest if $\Delta t$ is a typical CFL hydrodynamical
time step, it becomes unnecessary to treat the
detailed time evolution of $f$ at shocks with eq. [\ref{dceq}].  Instead
for GeV electrons we can assume immediately
downstream  of a shock that in response to diffusive shock
acceleration $f \propto p^{-q}$, where $q = min(q_s, q_o)$,
with $q_o = -~\partial \ln{f_o}/\partial \ln{p}$ the logarithmic
slope of the in-flowing electron distribution at momentum $p$, and
$q_s = 3r/(r-1)$, the standard diffusive shock acceleration power law
slope. In this $r = \rho_2/\rho_1$, the compression ratio through the
shock (\eg \cite{drury83}).

The above discussion also leads to the conclusion that, at least as it
emerges from shocks, the electron momentum distribution for
energies of direct relevance for radio emission will generally be
well-approximated by a power law. That emergent form may change
in subsequent propagation  over finite lengths due to 
radiative losses, mixing between particle populations with
different initial spectra or second-order Fermi acceleration. 
Nonetheless, this suggests an obvious simplification
to the energy distribution that greatly reduces the effort required
to solve eq. [\ref{dceq}] outside of shocks in RG flows.
In particular $q = -~\partial \ln{f_o}/\partial \ln{p}$
should be a slowly varying function of $\ln{p}$. Then, for
numerical purposes it makes sense to approximate the momentum
distribution as a piece-wise power law over bins of finite width 
in $\ln{p}$. The practical coarseness allowed for such a momentum grid 
will be determined by the expected curvature of the spectrum
and the information needed from the spectrum in the momentum range
of particular interest.

\subsubsection{Transport in Smooth RG Flows}

In smooth flows, eq. [\ref{dceq}] is directly
applicable. Here the
relative importance of advective and diffusive fluxes is given by
a ratio $u \delta x/\kappa$, where $u$ is a characteristic advection
speed across zone boundaries and $\delta x$ is the scale length of
$\nabla f$ in the smooth flow.
Assuming $\delta x > \Delta x$, it is easy to show with the
diffusive properties mentioned earlier, that on a typical
numerical grid with $\sim 10^3$ zones spanning the RG
dimensions in any direction, $u\delta x/\kappa > u\Delta x/\kappa >>1$. Thus,
advection of $f$ for low energy 
particles in the smooth 
flows should always dominate over spatial diffusion even if the spatial
diffusion coefficient is several orders of magnitude greater than
that we discussed next to the shocks. {\it We conclude that we can neglect the
spatial diffusion term in eq. [\ref{dceq}] for low energy electrons in
RGs except at shocks}.

\subsection{A Simple Electron Transport Scheme for RGs}

To solve eq. [\ref{dceq}] we should divide
a momentum range of interest into N bins bounded by $p_0,\ldots,p_N$.
It is most practical to work 
with logarithmic intervals in $p$,
so we generally use $y_i = \ln{p_i/p_0}$, giving bin widths, 
$\Delta y_i = y_{i+1} - y_i = \ln{p_{i+1}/p_i}$.
For a conventional, finite difference treatment of eq. [\ref{dceq}] it is
necessary to use bins with $\Delta y_i << 1$ (\eg \cite{kj91}).
However, with the conservative, finite volume approach defined here,
it is practical to use
much greater bin widths,  $\Delta y_i \gsim 1$, when the smooth
behavior of $\ln{f(p)}$ outlined in \S A.2 applies.

\subsubsection{Smooth Flows}

To begin, we can integrate equation [\ref{dceq}] within each momentum
bin to define $n_i = 4 \pi \int_{p_i}^{p_{i+1}} p^3 f d\ln{p}$ as the number of
electrons in the bin.
It is convenient (but not necessary) to normalize $n_i$ by the total plasma mass density, to
form $b_i = n_i/\rho$.
Then eq. [\ref{dceq}] becomes, for one spatial dimension the conservative,
finite volume equation,
\begin{equation}
{{d b_i}\over{d t}} = 4 \pi\left ( \frac{1}{3}\frac{\partial u}{\partial x}
+ \frac{D}{p^2}\frac{\partial \ln{f}}{\partial y} \right) \frac{p^3 f}
{\rho}
\left\vert_{_{p_{i}}}^{^{p_{i+1}}}\right.
+ \frac{\partial}{\partial x}
\left ( \langle\kappa_i\rangle \frac{\partial b_i}{\partial x}\right )
 +\frac{Q'}{\rho},
\label{simp1}
\end{equation}
where $<\kappa> =
{\int p^2 \kappa \nabla f dp \over \int p^2 \nabla f dp}$, $Q^\prime
= 4\pi  \int_{p_i}^{p_{i+1}} p^2~Q dp$, and $d~/dt$ is the
Lagrangian time derivative. 
No approximations have been added to obtain eq. [\ref{simp1}]
from eq. [\ref{dceq}]. 

Now we explicitly include the two simplifying features into eq. [\ref{simp1}]
that enable us to treat low energy electron transport efficiently in RGs.
We shall also define $Q'$ specifically, so that it accounts for
synchrotron losses and shock injection of low energy electrons.
Remember that eq. [\ref{dceq}] or [\ref{simp1}] applies only to the
smooth flows. 
In those regions we assume the piecewise power law
$f(p) = f_i~ (p/p_i)^{-q_i}$  within $p_i \le p \le p_{i+1}$, so that
\begin{equation}
n_i = 4 \pi {{f_i p_i^3} \over {q_i - 3}}
[1~-~({{p_i} \over {p_{i+1}}})^{q_i-3}].
\label{nieq}
\end{equation}
Since $f(p)$ is continuous we can also write $f_{i+1} = f_i (p_i/p_{i+1})^{q_i}$.
That enables us to find the fluxes at momentum boundaries in eq. [\ref{simp1}]
to account for
adiabatic expansion, second-order Fermi acceleration or radiative cooling.
Then also recall from \S A.2 that we can neglect the spatial
diffusive transport term in eq. [\ref{simp1}] for low energy
electrons within smooth portions of 
flows for the conditions expected in RG simulations.

Synchrotron aging of electrons is straightforward to include in this
formulation when there is rapid angular redistribution of electrons
(a necessary condition for the validity of eq. [\ref{dceq}]). 
Then at momentum $p$ there is  a flux $4\pi p^2 \dot p_s f$, with 
$\dot p_s = - \frac{p^2}{\tau_{so}\hat p}$, where $\tau_{so} = \frac{3}{4}
\frac{(m c)^2}{\sigma_T U_B}\frac{1}{\hat p}$ defines the cooling time at some
convenient momentum $\hat p$ (see, \eg \cite{jos74}).
In this expression, $\sigma_T$ is the Thomson cross section, 
and $U_B = B^2/(8\pi)$.
To account for inverse Compton cooling  from the cosmic microwave 
background $B^2$ should be modified to $B^2 + B^2_\mu$,
where $B_\mu = 3.2~\mu$G at the current epoch.
Putting these features together, we are left in smooth flows to solve 
the simple transport equation
\begin{equation}
{{d b_i}\over{d t}} = 4 \pi
\left (\frac{1}{3}\frac{\partial u}{\partial x}
-\frac{q D}{p^2}
+ \frac{1}{\tau_{so}}\frac{p}{\hat p}\right )
\frac{p^3 f} {\rho}
\left\vert_{_{p_{i}}}^{^{p_{i+1}}}\right..
\label{simple}
\end{equation}

Since eq. [\ref{simple}] is in conservation form, our main constraint is the accuracy
of the fluxes written on the right hand side. Given an initial set of $b_i$
and $q_i$, along with
underlying plasma properties, those can be found accurately at
the beginning of the time step, within the
piecewise power law approximation by using eq. [\ref{nieq}] and
the continuity of $f$. 
The fluxes can easily be time centered, by using the slope of
$f$, to achieve second order time accuracy in the update.
One must, of course, include a constraint on the time step, $\Delta t$,
beyond the MHD CFL condition. That can be expressed conservatively
to account for adiabatic, synchrotron and second-order acceleration
effects by the combined condition, 
$\Delta t < \max{(\Delta x/a, \tau_{so}\hat p/p_N, p^2/{(q D)_{min}})}$,
where $a$ is the fastest MHD signal speed.
The first condition, for adiabatic expansion, corresponds to
the standard CFL condition. The other two need to be
examined explicitly, but so long as the synchrotron cooling and 
second-order acceleration take place on ``hydrodynamical timescales'',
they will also usually be enforced by the CFL condition.

In general $q_i$ will also change during a time step. 
That set of values is
updated  from the new distribution $b_i$
by inverting eq. [\ref{nieq}], and again using the continuity
of $f$. An appropriate set of boundary conditions is also
required, of course. We expect spectral curvature will be common
at the highest momenta, but usually minimal at the lowest momenta.
So, for the simulations presented here we have
assumed that $q_i$ is continuous at $p_0$, but that the spectral
curvature is continuous at $p_N$. 
The solution to $q_i$ must be done iteratively.
As pointed out by \cite{jj98}, it is possible in principle to find $q_i$ exactly
if the momentum bins are of uniform size and one applies the
boundary condition $q_o = q_1$. For significantly curved spectra, however,
we have found a simple, approximate method to be relatively fast and
sufficiently accurate in these applications.
In this method we assume for the inversion of eq. [\ref{nieq}]
to find $q_i$, that $q$ is a linearly varying function of $y$
between the centers of adjacent momentum bins; that is, there is
uniform spectral curvature between adjacent bin centers. Thus,
$q_i$ is interpreted as the
average $q$ (or midpoint value) within the interval [$y_{i-1}, y_{i+1}]$.
Then it is straightforward to design
a Newton-Raphson iteration scheme to solve for each $q_i$. In practice
we use only a single iteration from the initial guess
\begin{equation}
q_i(trial) = 3 - {{\ln{\frac{b_{i+1}}{b_i}}}\over
{\Delta y_i}},
\end{equation}
which would be exact for a pure power law on uniform bins.

The number of bins necessary  in using eq. [\ref{simple}]
over a given momentum range
will depend on the amount of spectral curvature expected.
\cite{jj98} showed
if one can assume that $f(p)$ is a pure power law at any given
physical location, then two momentum bins are sufficient to follow the
evolution of $f$ accurately, accounting for adiabatic expansion,
advection and consequent mixing of different power laws.
In our application that is not adequate, since we may expect
significant deviations to develop from pure power law form.
On the other hand, from thermal injection energies to those $\sim$ GeV 
the relevant $\Delta\ln{p} < 20$, while a range $|\Delta q| = 2\Delta\alpha = 3$
would cover electron populations producing synchrotron spectra with
slopes $0.5 \le \alpha \le 2$, assuming $q \ge 4$. 
Greater curvature would involve spectral slopes, $q > 7$, so that
very few particles would be represented at the higher energies.
As demonstrated below we have found that it is
possible to capture the evolution of electron spectra 
utilizing only a few momentum bins for practical problems.
The computational cost of following each electron momentum bin is roughly
comparable to following each dynamical variable in the flow (of which
there are eight in MHD). The same
statement would apply for a conventional scheme to solve
eq. [\ref{dceq}], such as that we used in
\cite{kj91}; but there, a much larger number of momentum bins would
be necessary and much finer spatial resolution near shocks would
be needed to obtain meaningful solutions in RGs (see \S A.1 and
comments at the end of \S A.4).

\subsubsection{Shocks}

To model ``diffusive shock acceleration'', following the arguments of
\S A.1, we impose 
at detected shocks the analytic solution outlined in \S A.2.
That is, the emerging distribution has a form $f(p) \propto p^{-q}$,
where $q = 3r/(r-1)$, with $r$ the compression ratio through the shock
provided $q_s < q$, where $q$ is the in-flowing value in a momentum bin.
Normalization of $f$ is determined by the total number of electrons;
that being the sum of the incident population in the interval [$p_0,~p_N$]
and any injected 
electrons as defined below. 
We ignore shocks with density jumps $r < \frac{7}{4}$ ($q_s > 7; \alpha > 2$),
since they contribute very few nonthermal particles.

For fresh injection at shocks we adopt a commonly used simple model (\eg \cite{kj91});
namely, we inject a small, fixed fraction, $\epsilon$,
of the thermal electron flux through the shock. Thus inside a 
shock we solve the equation $d b/d t = Q'_{inj}/\rho$, where
$b = \sum b_i$ and
\begin{equation}
Q'_{inj} =  {{\epsilon W_s}\over{\mu_e m_H}}.
\end{equation}
$W_s$ is the Lagrangian shock velocity, $\mu_e$ is the electron
mean molecular weight, $m_H$ is the mass of the proton.
As for electrons flowing in from upstream, shock injected electrons are
distributed in momentum as a power law with the appropriate slope from
diffusive acceleration theory, as described before.

\subsection{Tests}

We have carried out numerous
tests of the method described above and find it to be both very
stable and to produce consistent results, provided roughly
$|\Delta q_{i,i+1}| < 0.75$. In the simulations described here we
have constrained $|\Delta q_{i,i+1}| < 0.67$.
Our various tests included 1-D adiabatic advection of analytically
defined convex and concave spectra, as well as shock tube
simulations. We carried out a variety of 2-D simulations, such as
spherical blast waves and 
obliquely intersecting shocks, where we could compare
analytic expectations of the resulting electron spectra.
We also tested the ability of the scheme to handle synchrotron aging by
following the evolution of an initial power law electron
distribution compared to the well-known analytic result of \cite{kard62}
(but, most commonly associated with \cite{jaf73})
for synchrotron aging with rapid pitch angle redistribution. 

A 2-D application of this scheme to the propagation of a young SNR 
impacting an interstellar cloud is presented in Jun \& Jones (1998),
and can serve to demonstrate its ability to capture expected
behaviors correctly for a spherical blast wave, as well as for mixed flows.
We offer here another simple 2-D test calculation that could be used
by other workers for quantitative comparison. It involves the interaction between
two obliquely intersecting plane hydrodynamical shocks, and is illustrated in Fig. 6.
Moderate strength shocks are more challenging to capture correctly with
regard to electron acceleration, so the example is designed to create
several shocks with Mach numbers in the range $1.5~-~10$.
There are five dynamical regions, as outlined below and as shown in
Fig. 6a.
Region 1 is at rest, with $\rho_1 = 1$ and $P_1 = 0.6$, giving unit sound
speed. Region 2 (also uniform) lies behind a Mach 10 right-facing shock 
initially placed vertically in the middle of the box. 
The initial conditions
in region 2 are, thus, $\rho_2 = 3.884$, $P_2 = 74.88$, $u_{x2} = 7.425$,
$u_{y2} = 0$. Region 5 similarly lies behind a Mach 4 shock, which, in
this case is propagating diagonally up and to the left. This shock
was initially placed so that it bisected both boundaries,
and, thus, was just touching the vertical shock. Initial conditions in 
region 5 were, $\rho_5 = 3.368$, $P_5 = 11.85$, $u_{x5} = -1.989$ and
$u_{y5} = 1.989$.

Propagation of each of the two original shocks into the postshock flows of
the other creates new regions 3 and 4, which are separated by a slip
line. We carried out this test on a $512\times 512$ Cartesian grid.
Except for the right edge of the box standard continuous boundaries were
used. On the right these were modified to include the symmetry of the
flow there, thus preserving its self-similarity.
The whole flow pattern is self-similar, in fact, except near the
bottom of the box, where it is not possible to construct suitable
boundary conditions. That leads to the curvature in the shocks
visible there, and the incipient vortex at the bottom of the slip line.
Using shock polars (\eg, Courant \& Friedrichs (1976)) it is simple
to compute that regions 2 and 3 are separated by a Mach 1.51
shock, while regions 4 and 5 are divided by a shock with Mach number 3.66.

For this particular test there was an initial cosmic-ray electron
population that was a uniform 1\% of the thermal electron population
in each region. The cosmic-rays arbitrarily were given an initial, uniform power-law form with
index, $q = 4.5$, corresponding to original injection at a Mach 3 shock.
Many other spectral choices could be made, as well.
There was no injection of cosmic-ray electrons
at the shocks during the simulated flow from this point, 
although it is straightforward to include that effect, as well. 

Fig. 6b illustrates the computed distribution of spectral indices after
the flow has evolved. Since the flow is self-similar, except for the
mentioned boundary effects, the chosen time does not matter, so long
as it is sufficient to allow significant material to pass through
each shock. As mentioned
the electrons initially had the same spectrum everywhere. Particle
acceleration at the shocks during their propagation during the simulation
has modified the electron spectra in regions 2', 3'', 4' and 5'. The steady-state
spectrum behind a Mach 1.51 shock (region 3) from a monoenergetic
incident flux would be $q = 7.11$. Since
that is steeper than the initial spectrum there is no change between
regions 2 and 3'. However, region 3'' was first processed by the Mach 10
shock, so we find, as expected, electrons there to have the same form
as those in region 2'.
From the analytical jump conditions we predict the spectral indices in
regions 2', 4', and 5' to be 
respectively, $q_{2'} = 4.040$, $q_{4'} = 4.323$, $q_{5'} = 4.267$.
In this numerical example they come out to be $q_{2'} = 4.041$, $q_{4'} = 4.333$,
and $q_{5'} = 4.299$, with variations less than or of order $10^{-4}$.
Thus, the errors in $q$, which are naturally greatest for
oblique shocks on a Cartesian grid, are all less than 1\%.
Since there is little mixing and no synchrotron aging during this
test calculation, all the spectra should very good power laws, which
we confirmed using 4 momentum bins covering $\Delta \ln p = 20$.

This test also demonstrates the ability of our routine to advect the
electrons cleanly in the different momentum bins, since the spectral 
indices are computed from the
actual particle distributions among the distinct momentum bins. So, for
example, the population differences  between regions 3' and 3''
in the higher momentum bins are more than an order of magnitude,
and that difference is maintained accurately with a transition over
about three spatial zones. Any error or significant diffusion there 
would lead to a spurious flattening of the spectrum along the boundary 
region 3'. The spectra there computed for the two highest 
momentum bins is almost precisely the same as for the two lowest bins.
Similar comments apply to the slip line between regions 3'' and
4'. Recall, of course, that we assume for these calculations that real, 
physical diffusion is very small on scales of the computational zones.

To demonstrate further the behavior of our electron transport scheme we include 
Fig. 7, which shows a snapshot of the results from three 2D test simulations exactly like 
the {\bf Model 3} jet described in \S 3, except that the spatial
resolution here was only half as fine. In these calculations an 
electron population with $p_0 = $mc, $p_N \approx 10^5$ mc
and a power law index $q = 4.4$ 
enters with the jet on the left of the grid. That population is modified
by adiabatic and synchrotron cooling and by any shocks
encountered along the way. The grayscale image shows the log of the
total electron number near the end of the simulation. For
numerical convenience, the computation also includes a small population of
steep spectrum ($q = 7$) background electrons in the ambient
medium. They are faintly visible
in the image, especially through their compression at the jet bow shock.
Fig. 7 also shows the computed electron momentum spectral
index at four selected locations on the grid. Each plot gives the values of
$q_i$ found in simulations with the number of momentum bins, $N$ set
to $4$ (triangles), $8$ (stars) and $12$ (squares). 

Note first the
plot (b) in Fig. 7, which shows the spectrum of the background
population, changed only in response to synchrotron aging in a uniform
background magnetic field.
The solid curve gives the prediction,
$q(p,t) = - \partial \ln{f}/\partial\ln{p}$
from \cite{kard62}; namely,
\begin{equation}
q(p,t) = q_0 + (q_0 - 4)\frac{\frac{t~p}{\tau_{so}\hat p}}{1~-~\frac{t~p}{\tau_{so}\hat p}},
\label{slope}
\end{equation}
where $q_0$ is the initial slope. In this case $q_0 = 7,$
and the time shown was $t = \tau_{so}$.
Equation \ref{slope} is valid, of course, only for
$\frac{t~p}{\tau_{so}\hat p} < 1$, since higher momentum particles have all
cooled below $p_{cool} = \hat p \frac{\tau_{so}}{t}$.
In this case $p_{cool} = \hat p = 10^4$ mc.
All three numerical results agree with eq. [\ref{slope}] in the range of
momenta where the curvature is less than the numerical imposed constraint mentioned
earlier; namely that the change in $q_i$ between adjacent bins is less 
than $\frac{2}{3}$.
For $N = 4$ violation occurs above $p \sim 10^{3}$ mc in this case, but for
both $N = 8$ and $N = 12$ the numerical results are in reasonable
agreement below about $\frac{1}{3} p_{cool}$.
The deviation at higher momenta has little practical consequence, since
the numerical slope there is still so steep and the associated $b_i$ so
small that it will have little influence on the evolution at lower
momenta (vanishing flux contribution) and would contribute
negligibly to emission processes. 

Similarly with each of the other three particle spectra shown (all corresponding
to the dynamical time displayed in the image), the agreement between
the $N = 8$ and $N = 12$ solutions is very good until the computed 
spectra are too steep to influence observable properties. For the
two spectra (c) and (d, where curvature is small below 
$p \sim 10^4$ mc,
the $N = 4$ solution agrees quite well with the other two. We conclude that
the scheme is effective and reasonably robust. When there is only modest
spectral curvature in the momentum range of interest, as few as 4
momentum bins may be adequate to capture the evolution of the
spectrum. For the first two RG models presented in \S 3,
synchrotron aging is negligible, so we use $N = 4$ in
those calculations. In situations where strong curvature needs to be captured,
such as those involving strong synchrotron aging ,
a larger number of bins are needed. In the {\bf Model 3} presented in this paper 
we consider emission almost entirely from electrons  having
$p < 10^4$ mc, and we judge there that $N = 8$ is adequate for the issues 
under discussion.
This ability to capture basic behaviors of electron transport in RG flows
with only a few momentum zones results from the conservative nature of
the scheme, the quasi-power law properties of the electron distribution
and the fact that the microphysics of the electron transport can
be analytically well-approximated in smooth flows and at shocks. The impact of
this fortuitous outcome is clear when one considers that any scheme
to follow the electrons will require a computational effort per
variable tracked that is comparable to the effort necessary to follow
a dynamical variable, such as the gas density. In this scheme we
need $ 2 N$ variables ($b_i,~q_i$), with $N \sim$ a few, so that the
net effort to follow the relativistic electrons is comparable
to the net MHD effort in the calculation. In our experience, direct
solution of eq. [\ref{dceq}] requires a momentum resolution $p_{i+1}/p_i \lsim 1.05$,
so we would need $\sim 250$ momentum bins (to say nothing of the
spatial resolution issues emphasized earlier). Thus, the cost of
following the electrons would exceed the cost of the MHD by at least
an order of magnitude. Since these are computationally expensive
simulations anyway, especially in 3D, that would make ordinary
schemes completely impractical at present.

\clearpage

\begin{center}
{\bf FIGURE CAPTIONS}
\end{center}
\begin{description}

\item[Fig.~1] 
{Dynamical flow properties of the simulated MHD jet at $t = 10.67$.
Normal gray scale is used, with high tones representing high values.
Shown top to bottom are: (a) Log gas density, (b) Log magnetic pressure,
(c) Gas compression, with shocks showing high tones.}

\item[Fig.~2] 
{(a,b): Log synchrotron emissivity computed at 1.4 GHz from
the simulated electron distributions for {\bf Models 1} and {\bf 2} 
at the same time and with the same MHD properties as in Fig. 1.
(c): ``Pseudo emissivity'', $j_c$, using only MHD variables.
The same dynamic range is shown for all; namely, $3\times 10^3$
below the maximum value in each case.}

\item[Fig.~3]
{Synchrotron emissivity spectral index distributions, $\alpha_1$
and $\alpha_2$, for the
electron acceleration {\bf Models 1} and {\bf 2}, as shown in Fig. 2.
(a) Model 1, (b) Model 2. 
Regions are included only if they show in
Fig. 2 and have $\alpha \le 0.8$. }

\item[Fig.~4]
{(a): Log synchrotron emissivity at 1.4 GHz computed from {\bf Model 3},
which includes strong synchrotron aging effects. The dynamic range is
increased over Fig. 2 by a factor 10 to $3\times 10^4$ to capture the
same physical regions. (b): Synchrotron emissivity spectral index
distribution, $\alpha_3$ for {\bf Model 3}. 
Regions are included only if they show in Fig. 4a.
(c): Synchrotron spectral curvature between 1.4 GHz and 5 GHz for
the same regions shown in Fig. 4b.}

\item[Fig.~5]
{Scatter plot displaying the distribution of 
spectral index vs emissivity 
[(a), (c), (e)] and spectral magnetic field vs index [(b), (d), (f)]
values for {\bf Models 1, 2} and {\bf 3}
at $t = 10.67$
Each point represents one zone on the
simulation grid, using a stride of 2 to reduce the size of the 
plot file. The number of points in a given range on these plots
can be seen as the area on the computational grid with those properties.
The magnetic field is normalized by $B_{background}$.} 

\item[Fig.~6]
{A 2-D test calculation of the relativistic electron transport scheme
described in section A.3. It involves the evolution of two obliquely
intersecting shocks, as described fully in the text. The computational
box was given dimensions $1.5\times 1.5$ and the time shown is $t = 0.05$.
(a) Gas density, with the five dynamical flow regions labeled. High tones
are high densities.
(b) Electron momentum spectral index. High tones are flatter spectra.}

\item[Fig.~7]
{Tests of the electron transport scheme based on simulations identical
to {\bf Model 3} (see Table \ref{tabmod}), except these were done
at half the spatial resolution. Gray scale image: Log of the relativistic
electron density. (a),$\ldots$,(d): Computed electron momentum
distributions at indicated locations using 4 (triangle), 8 (star), 
and 12 (square) momentum bins to cover the range shown. The plot (b)
was taken from a region subject only to synchrotron aging effects, and
shows the results when $t = \tau_{so}$ (see eq. [\ref{tcool}], with
$\hat p = 10^4$, along with the analytic solution (solid curve).}

\end{description}

\end{document}